\documentclass[12pt,aps]{article}
\usepackage{amssymb}
\usepackage{amsmath}
\newcommand{\ket}[1]{{|#1\rangle}}
\newcommand{\bra}[1]{{\langle#1|}}

\newcommand\be{\begin{equation}}
\newcommand\ee{\end{equation}}
\newcommand\bea{\begin{eqnarray}}
\newcommand\eea{\end{eqnarray}}

\newcommand\bk[2]{\langle #1|#2\rangle}
\newcommand\pty{{$\cal PT$}-symmetry}
\newcommand\ptc{{$\cal PT$}-symmetric}

\newcommand\pte{{$\cal PT$}-invariance}
\begin{document}
\noindent{\Large \sf The Gram Matrix of a {\cal PT}-symmetric Quantum System}\\ \\

\noindent{\sc \normalsize Stefan Weigert} \\ \\
\noindent{\normalsize \rm $\bra{\mbox{Hu}}\mbox{MP}\rangle$ -
\rm Department of Mathematics, University of Hull\\
UK-Hull HU6 7RX, United Kingdom \\
{\tt S.Weigert@hull.ac.uk}\\
October 2003 \\ \\

\parbox[height]{13cm}{\small {\bf Abstract}:
The eigenstates of a diagonalizable \ptc\ Hamiltonian satisfy
unconventional completeness and orthonormality relations. These
relations reflect the properties of a pair of bi-orthonormal bases
associated with non-hermitean diagonalizable operators. In a
similar vein, such a dual pair of bases is shown to possess, in
the presence of \pty , a Gram matrix of a particular structure:
its inverse is obtained by simply swapping the signs of some its
matrix elements.}
\\ \\

The spectrum of a non-hermitean Hamiltonian $\hat H$ is {\em real}
\cite{spectra} if the Hamiltonian is invariant under the combined
action of self-adjoint parity $\cal P$ and time reversal $\cal T$,
\be
[ \hat H , {\cal P T} ] = 0 \, ,
\label{ptsymm} \ee
and if the energy eigenstates are invariant under the operator
${\cal PT}$. Pairs of {\em complex conjugate} eigenvalues are also
compatible with \pty\ but the eigenstates of $\hat H$ are no
longer invariant under $\cal PT$. Wigner's representation theory
of anti-linear operators \cite{wigner60}, when applied to the
operator ${\cal PT}$ \cite{weigert02}, explains these observations
in a group-theoretical framework. Alternatively, they follow from
the properties of {\em pseudo-hermitean} operators
\cite{multimostafazadeh} satisfying $\eta \hat H \eta^{-1} = {\hat
H}^\dagger$ equivalent to Eq. (\ref{ptsymm}) if $\eta = {\cal P}$.

Consider a (diagonalizable) {\em non-hermitean} Hamiltonian $\hat
H$ with a discrete spectrum \cite{wong67}. The operators $\hat H$
and and its adjoint $\hat H^\dagger$ have complete sets of
eigenstates:
\be \label{nonhermH}
\hat H \ket{E_n} = E_n \ket{E_n} \, , \quad
 \hat H^\dagger \ket{E^n} = E^n \ket{E^n} \, , \quad n= 1,2, \dots \, ,
\ee
with, in general, complex conjugate eigenvalues, $E^n = E_n^*$.
The eigenstates constitute {\em bi-orthonormal} bases in $\cal H$
with two resolutions of unity,
\be \label{completeness}
\sum_n \ket{E^n}\bra{E_n}
 = \sum_n \ket{E_n}\bra{E^n}
 = \hat I \, ,
\ee
and as dual bases, they satisfy orthonormality relations,
\be\label{dual}
\bra{E^n} E_m \rangle = \bra{E_m} E^n \rangle = \delta_{nm} \, ,
\quad m,n = 1,2, \ldots
\ee

It has been shown \cite{weigert03-1} that \pty\ of the Hamiltonian
(\ref{nonhermH}) implies the existence of a simple relation
between the state $\ket{E_n}$ and its dual partner $\ket{E^n}$,
\be \label{newprop}
 \ket{E^n} = s_n {\cal P} \ket{E_n}
           = {\cal P} {\cal C}_s\ket{E_n} \, , \quad
 s_n = \pm 1 \, ,
\ee
where the {\em signature} $s = (s_1,s_2, \ldots)$ depends on the
actual Hamiltonian, and the operator ${\cal C}_s$ is given by
\be
{\cal C}_s =  \sum_m s_m \ket{E_m}\bra{E^m}
           \neq \sum_m s_m \ket{E^m}\bra{E_m}
           = {\cal C}_s^\dagger \, .
\ee

 The unconventional
completeness and orthogonality relations which are characteristic
for \ptc\ systems having real eigenvalues only are a direct
consequence of Eq. (\ref{newprop}). Numerical work suggests
\cite{bender+02} that there is a completeness relation of the form
\be
 \sum_n s_n \phi_n (x) \phi_n(y) =\delta (x-y) \, ,
\label{unconcompl} \ee
which is a consequence of the completeness relations
(\ref{completeness}),
\be \label{newcompleteness}
 \sum_n \ket{E_n}\bra{E^n}
     = \sum_n s_n \ket{E_n}\bra{E_n} {\cal P} = \hat I \, ,
\ee
when rewritten (in the position representation) by means of Eq.
(\ref{newprop}).

 Similarly, the orthonormality condition for dual
states turns into a relation which has been interpreted
\cite{japaridze02} as the existence of a non-positive scalar
product among the eigenstates of $\hat H$. To see this, write the
scalar product (\ref{dual}) in the position representation, using
again (\ref{newprop}) and \pte ,
\be
\bk{E^n}{E_m} =s_n \bra{E_n} {\cal P} \ket{E_m}
              = s_n \int dx \, \phi_n (x) \phi_m (x)
              = \delta_{nm} \, ,
\label{scprtrf} \ee
or $(\phi_n,\phi_m) = s_n \delta_{nm}$, in the notation of
\cite{bender+02}.

Let us now turn to the properties of the {\em Gram} matrix ${\sf
G}$ of a \ptc\ quantum system. For a general bi-orthonormal pair
of bases one defines the Gram matrix by
\be \label{gram}
{\sf G}_{mn} = \bra{E_m} E_n \rangle \, ;
\ee
its inverse ${\sf G}^{-1}$ exists since the states $\{ \ket{E_m}
\} $ are linearly independent, and its matrix elements are given
by
\be \label{gram-1}
\left({\sf G}^{-1}\right)_{mn} = \bra{E^m} E^n \rangle \equiv {\sf
G}^{mn} \, .
\ee
Given the states $\{ \ket{E_m} \}$ and hence ${\sf G}$, one finds
the {\em dual} states $\{ \ket{E^n} \}$ through the inversion of
${\sf G}$:
\be \label{godual}
\ket{E^n} = \sum_m \ket{E_m}\bra{E^m} E^n \rangle
          = \sum_m {\sf G}^{mn} \ket{E_m}
          \equiv \sum_m \left({\sf G}^{-1}\right)_{mn} \ket{E_m} \, .
\ee

Eq. (\ref{newprop}) establishes a simple link between each state
$\ket{E_m}$ and its partner $\ket{E^m}$ which will be shown now to
imply a simple relation between ${\sf G}$ and its inverse,
\be \label{simpleinv}
{\sf G}^{-1} = {\sf S} {\sf G} {\sf S} \, , \qquad \mbox{where}
\quad {\sf S} = \mbox{diag} (s_1, s_2, \ldots) \, ,
\ee
with ${\sf S}$ a real diagonal matrix, being determined entirely
by the signature $s$ of the system studied. To derive this
relation, multiply the resolutions of unity given in
(\ref{completeness}) with each other,
\be \label{multicompleteness}
\hat I = \left(\sum_m \ket{E^m}\bra{E_m}\right)
         \left(\sum_n \ket{E_n}\bra{E^n}\right)
 = \sum _{m,n} {\sf G}_{mn} \ket{E^m}\bra{E^n} \, ,
\ee
and use Eq.(\ref{newprop}) giving
\be
\hat I = \sum _{m,n} {\sf G}_{mn} s_m {\cal P} \ket{E_m}                                          \bra{E_n} {\cal P} s_n \, .
\ee
Finally, multiply this equation with $\bra{E^k}{\cal P}$ from
the left and with ${\cal P}\ket{E^l}$ from the right to find
\be \label{meinverse}
{\sf G}^{kl} \equiv \left({\sf G}^{-1} \right)_{kl}
  = \sum _{m,n} {\sf G}_{mn} s_m \delta_{km} s_n \delta_{nl}
  = s_k {\sf G}_{kl} s_l \, ,
\ee
which is the matrix version of Eq. (\ref{simpleinv}).

As a result, the inverse ${\sf G}^{-1}$ of the Gram matrix ${\sf
G}$ is obtained by multiplying each of the matrix elements ${\sf
G}_{mn}$ by the product $s_m s_n$ which takes the values $\pm 1$
only. Due to $s_m^2=1$, the diagonal elements of the Gram matrix
and those of its inverse are necessarily equal. Furthermore,
having determined the eigenstates $\ket{E_m}$ of a \ptc\
Hamiltonian operator $\hat H$ and hence its Gram matrix via
$\bra{E_m} E_n \rangle$, the dual states are given by
\be \label{godual2}
\ket{E^n} = \sum_{m,n} s_m s_n {\sf G}_{mn} \ket{E_m} \, ,
\ee
thus considerably simplifying Eq. (\ref{godual}): the usually
cumbersome inversion of ${\sf G}$ can be avoided.

\end{document}